\documentclass[10pt,journal,compsoc]{IEEEtran}

\ifCLASSOPTIONcompsoc
  \usepackage[nocompress]{cite}
\else
  \usepackage{cite}
\fi

\usepackage[dvips]{graphicx}

\ifCLASSINFOpdf
\else
\fi

\begin{document}
%
% paper title
% Titles are generally capitalized except for words such as a, an, and, as,
% at, but, by, for, in, nor, of, on, or, the, to and up, which are usually
% not capitalized unless they are the first or last word of the title.
% Linebreaks \\ can be used within to get better formatting as desired.
% Do not put math or special symbols in the title.
\title{Learning, Visualizing, and Exploiting a Model for the Intrinsic Value of a Batted Ball}

\author{Glenn~Healey,~\IEEEmembership{Fellow,~IEEE}
\IEEEcompsocitemizethanks{\IEEEcompsocthanksitem G. Healey is with the Department
of Electrical Engineering and Computer Science, University of California, Irvine,
CA, 92617.\protect\\
% note need leading \protect in front of \\ to get a newline within \thanks as
% \\ is fragile and will error, could use \hfil\break instead.
E-mail: ghealey@uci.edu
%\IEEEcompsocthanksitem }% <-this % stops a space
}
\thanks{Manuscript received February 19, 2016;}}

\IEEEtitleabstractindextext{%
\begin{abstract}
We present an algorithm for learning the intrinsic value of a batted ball in baseball.
This work addresses the fundamental problem of separating the value of a batted ball
at contact from factors such as the defense, weather, and ballpark that can affect
its observed outcome.  The algorithm uses a Bayesian model to construct
a continuous mapping from a vector of batted ball parameters to an intrinsic measure
defined as the expected value of a linear weights representation for run value.
A kernel method is used to build nonparametric estimates for the component 
probability density
functions in Bayes theorem from a set of over one hundred thousand batted ball measurements
recorded by the HITf/x system during the 2014 major league baseball (MLB) season. 
Cross-validation is used to determine the optimal vector of smoothing parameters for the
density estimates.  Properties of the mapping are visualized by considering reduced-dimension
subsets of the batted ball parameter space.  We use the mapping to 
derive statistics for intrinsic quality of contact for
batters and pitchers which have the potential to improve the accuracy of player models and forecasting systems.
We also show that the new approach leads to a simple automated measure of contact-adjusted
defense and provides insight into the impact of environmental variables on batted balls.
\end{abstract}

% Note that keywords are not normally used for peerreview papers.
}

% make the title area
\maketitle

\IEEEdisplaynontitleabstractindextext

\IEEEpeerreviewmaketitle

\ifCLASSOPTIONcompsoc
\IEEEraisesectionheading{\section{Introduction}\label{sec:introduction}}
\else
\section{Introduction}
\label{sec:introduction}
\fi

\IEEEPARstart{T}{he} 
success of a major league baseball team depends on its ability to
predict the future performance of players.   This has led to the development of 
forecasting systems that can inform personnel
decisions which routinely result in player contracts worth tens of millions
of dollars.
Most forecasting systems are based on a process that estimates a player's current
talent level and another process that predicts how that talent level will change in the 
future~\cite{Silver}.
The first process generates a set of 
statistics that represent various player 
attributes using weighted averages of past observations.
Each statistic is then regressed to the mean by an amount that depends on the reliability 
of the statistic and the sample size~\cite{Stein} \cite{Tangobook}.
The second process utilizes a model for how each statistic changes as a 
player ages.  While
most statistics tend to improve for young players and decline for older players
there are significant differences in the aging curves for different skills 
\cite{Bradburyaging}.  Forecasting systems may also account for contextual 
variables such as a player's home ballpark during generation of the current 
talent estimate and the future projections \cite{pecota}.

Statistics for batters and pitchers that depend on the fate of 
batted balls tend to have a lower reliability than statistics 
that do not \cite{Carleton1} \cite{Carleton2}.
This occurs because a number of 
variables such as the response time of fielders, the texture of the infield
grass, and the ambient weather conditions 
contribute variation to statistics like batting average that depend on the outcome of batted 
balls.   Other statistics such as strikeout rate are less sensitive to these 
sources of variability and, as a result, provide a higher reliability.
Unsurprisingly,  the prediction of a player's future results on batted balls 
is often cited as the most challenging problem for a forecasting 
system~\cite{Baumer}\cite{Dupaul}.   
Since about 70 percent of major league plate 
appearances result in a batted ball, an effective approach for addressing
this challenge is of critical importance to a system's utility.

In this paper we develop a method for assigning an intrinsic value to batted balls
at contact.  This approach separates the intrinsic value of a batted 
ball from its outcome and, in the process, removes the confounding effects of
factors such as 
the defense, the weather, the ballpark, and random luck.
As a result, we are able to define batted ball statistics 
for batters and pitchers that are less subject to random variation
than statistics that are based on batted ball outcomes.
The new statistics have the additional advantage 
of separating components of a 
player's value that are intermingled using traditional statistics.   Hitter
descriptors such as batting average and slugging percentage, for example, are 
influenced by a player's running speed in addition to his batting ability since
faster runners are more likely to beat out infield hits or stretch singles into
 doubles.  With the new approach, a model for a player's offensive value 
can include a
statistic that captures the intrinsic value of his batted balls and another statistic
that captures his running speed. 
Similarly, a model for a pitcher's value can include a statistic for the
intrinsic value of opponent batted balls and another statistic for the
pitcher's fielding ability.
The generation of separate statistics to measure distinct skills benefits a forecasting
system because these statistics may be regressed and projected individually using
their specific reliability values and aging curves.

The model for the intrinsic value of a batted ball is derived from HITf/x data \cite{Jensen}
provided by Sportvision.  The HITf/x system uses multicamera video data to estimate
the three-dimensional speed and direction of batted balls after contact.
Using HITf/x measurements for more than one hundred thousand batted balls from the 2014 season, 
we construct a continuous mapping from the batted ball parameters
to intrinsic value as defined by the expected weighted on base 
average (wOBA)~\cite{Tangobook}.  The mapping is learned using a Bayesian model
that employs a kernel method to generate nonparametric estimates for 
the component probability density functions.
A cross-validation scheme is used to learn the optimal vector of smoothing bandwidths
for the batted ball parameters.   We show that the mapping has a significant
dependence on the handedness of the batter which leads to the generation
of separate functions for left-handed and right-handed batters.  
We use the mapping to define statistics that measure intrinsic contact quality for
batters and pitchers.
We also show that the mapping leads to a simple automated technique for measuring 
contact-adjusted team defense which could serve as a starting point for a 
HITf/x-based defensive metric.  The analysis also provides insight into
the impact of environmental factors on the outcome of a batted ball.

\vspace{-0.1cm} \section{HITf/x Data}

HITf/x is a system developed by Sportvision that uses image sequences acquired by
two cameras to estimate the initial trajectory of a batted ball in three dimensions.
The system uses the estimated trajectory to derive several batted ball descriptors. 
The speed $s$ is an estimate of the ball's initial speed in
three dimensions. 
The vertical launch angle~$v$ is the angle that the batted ball's 
initial velocity vector makes with the plane of the playing field where a vertical angle
of $-90^{\circ}$ is straight down and a vertical angle of $+90^{\circ}$ is straight up.
The horizontal angle $h$ specifies the direction of the projection of the batted
ball's initial velocity vector onto the plane of the playing field where  
the direction toward first base has a horizontal angle of $45^\circ$
and the direction toward third base has a horizontal angle of $135^\circ.$
The speed $s$ is reported in miles per hour and 
the angles $v$ and $h$ are reported in degrees.

\begin{figure}
\begin{center}
\begin{tabular}{c}
\vspace*{0.3cm}
\hspace*{-0.8cm}\includegraphics[height=7.48cm,width=11.22cm]{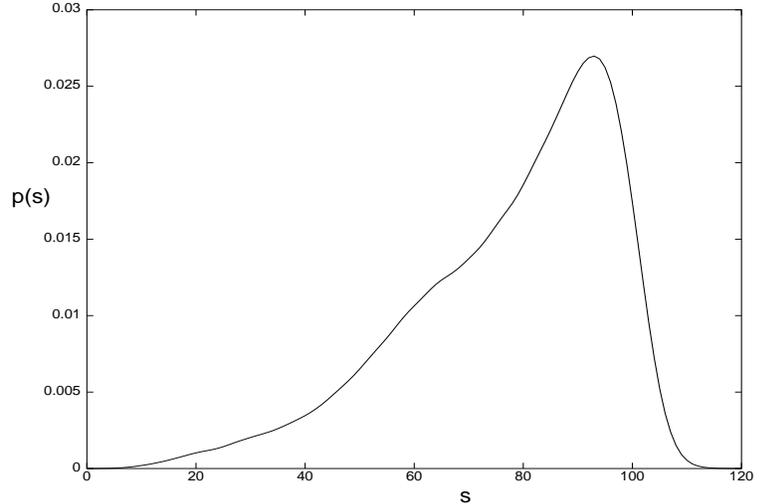}
\end{tabular}
\end{center}
\vspace*{-0.6cm}\caption{Distribution of initial speeds (mph) for batted balls in 2014}
{ \label{speedhist}
}
\end{figure}

\begin{figure}
\begin{center}
\begin{tabular}{c}
\vspace*{0.3cm}
\hspace*{-0.8cm}\includegraphics[height=7.48cm,width=11.22cm]{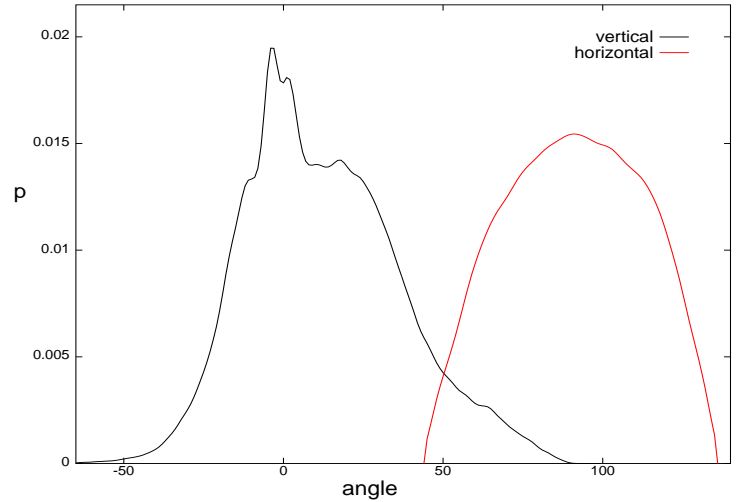}
\end{tabular}
\end{center}
\vspace*{-0.6cm}\caption{Distribution of vertical and horizontal angles (degrees)}
{ \label{vhhist}
}
\end{figure}

Figures~\ref{speedhist} and \ref{vhhist} plot the distribution 
for $s,v,$ and
$h$ for batted balls hit during the 2014 MLB regular season 
having horizontal angles in fair territory
$(h \in [45^\circ,135^\circ])$ after excluding bunts.  We see that the peak
of the speed distribution occurs at about 93 mph and that the peaks of
the vertical and horizontal angle distributions occur at about zero and 
ninety degrees respectively.
The result of a batted ball has a strong dependence on the $(s,v,h)$
parameter vector.  For example, a 
vector of $(75, 70,55)$ 
is typically a pop up to the first baseman, a vector of $(60, -10,105)$ is 
typically a ground ball
to shortstop, and a vector of $(100, 25,125)$ usually results in 
a home run to left field.   
HITf/x data was quickly shown to provide  
significant advantages for analysis over previous data that included only
a ground ball, line drive, or fly ball descriptor for each batted 
ball \cite{Cartwright}.
Early HITf/x studies \cite{Fast111611} \cite{Fast112211} also demonstrated 
that both batters and pitchers have some control over their average batted ball 
speed in the plane of the playing field and that this speed is correlated 
with the batted ball outcome.

\section{Learning the Value of a Batted Ball}
\label{sec-learning}

\subsection{Bayesian Foundation}
\label{sec-bayes}

Given a set of observed batted balls 
and their outcomes, we will develop a method for learning 
the dependence of a batted ball's value on its
measured parameters.
Let $x_i = (s_i,v_i,h_i)$ for $i = 1, 2, \dots, n$ be a set of $n$  
observed batted ball vectors where each vector has an associated
outcome such as a single, a double, or an out. 
Using Bayes theorem, the probability of an outcome $R_j$ given a 
measured vector  $x = (s,v,h)$ is given by 

\begin{equation}
P(R_j | x) = { {p(x | R_j) P(R_j) } \over { p(x) }}
\label{eq:bayes}
\end{equation}

\noindent where $p(x | R_j)$  is the conditional probability density
function for $x$ given outcome $R_j,$ $P(R_j)$ is the prior probability of 
outcome $R_j,$ and $p(x)$ is the probability density function 
for~$x.$   Linear combinations of the $P(R_j | x)$ 
probabilities for different outcomes can be used to model the expected value of
statistics such as batting average, wOBA, 
and slugging percentage as a function of the batted ball vector $x.$  
For a given batted ball, therefore, 
these statistics provide a measure of 
value that is separate from the batted ball's particular outcome.

\subsection{Kernel Density Estimation}
\label{sec-kde}

The goal of density estimation for our application 
is to recover the underlying 
probability density functions $p(x | R_j)$ and $p(x)$ 
in equation~(\ref{eq:bayes}) from the set
of observed batted ball vectors $x_i = (s_i,v_i,h_i)$ and their outcomes.
Given the typical positioning of defenders on a baseball field and the 
various ways that an outcome such as a single can occur, we expect a
conditional density $p(x | R_j)$ to have a complicated
multimodal structure.  Thus, we use a nonparametric technique
for density estimation.

We first consider the task of estimating $p(x)$ from the $n$ batted ball
vectors $x_i.$  
Kernel methods \cite{Sheather} 
which are also known as Parzen-Rosenblatt \cite{Parzen} \cite{Rosenblatt} window methods
are widely used for nonparametric density estimation.
A kernel density estimate for $p(x)$ is given by

\begin{equation}
\widehat p(x) = {{1} \over {n}} \sum_{i=1}^{n} K(x - x_i)
\label{eq:kde}
\end{equation}

\noindent where $K(\cdot)$ is a kernel probability density function that is
typically unimodal and centered at zero.  
A standard kernel 
for approximating a $d-$dimensional density is the zero-mean Gaussian 

\begin{equation}
K(x) = {{1} \over {(2\pi)^{d/2} |\Sigma|^{1/2}}} \exp \left [{  - {{1} \over {2}} x^T \Sigma^{-1} x} \right ]
\label{eq:gaussian}
\end{equation}

\noindent where $\Sigma$ is the $d \times d$ covariance matrix.  For this kernel, 
$\widehat p(x)$ at any $x$ is the average of a sum of Gaussians centered at the sample
points $x_i$ and the covariance matrix $\Sigma$ determines the amount and orientation
of the smoothing.   $\Sigma$ is often chosen to be the product of a scalar 
and an
identity matrix which results in equal smoothing in every direction.  However,
we see from figures~\ref{speedhist} and \ref{vhhist} 
that the distribution for $v$ has 
detailed structure while the distributions for $s$ and $h$ are 
significantly smoother.  Thus, to
recover an accurate approximation $\widehat p(x)$ the covariance matrix
should allow different amounts of smoothing in different directions.
We enable this goal while also reducing the number of unknown parameters by adopting
a diagonal model for~$\Sigma$ with variance 
elements $(\sigma_s^2, \sigma_v^2, \sigma_h^2).$  For our 
three-dimensional data, this allows $K(x)$ to be written as a product of three
one-dimensional
Gaussians 

\begin{equation}
K(x) = {{1} \over {(2\pi)^{3/2} \sigma_s \sigma_v \sigma_h }}  \exp \left [ {  - {{1} \over {2}}} 
\left ( {{s^2} \over {\sigma_s^2}} + {{v^2} \over {\sigma_v^2}} + {{h^2} \over {\sigma_h^2}} \right )   \right ]
\label{eq:oneD}
\end{equation}

\noindent which depends on the three unknown bandwidth parameters 
$\sigma_s, \sigma_v,$ and $\sigma_h.$

\subsection{Bandwidth Selection}
\label{sec-bandsel}

The accuracy of the kernel density estimate $\widehat p(x)$ is highly dependent
on the choice of the bandwidth vector $\sigma = (\sigma_s,\sigma_v,\sigma_h)$ 
\cite{DudaHartnew}.
The recovered $\widehat p(x)$ will be spiky for small values of the 
parameters and, in the limit, will tend to a sum of Dirac delta functions
centered at the $x_i$ data points as the bandwidths approach zero.  Large bandwidths,
on the other hand, can induce excessive smoothing which causes the loss
of important structure in the estimate of~$p(x).$  A number of 
bandwidth selection techniques 
have been proposed and a recent survey of methods and software
is given in \cite{keddR}.
Many of these techniques are based on 
maximum likelihood estimates for $p(x)$ which select $\sigma$ so that
$\widehat p(x)$ maximizes the likelihood of the observed 
$x_i$ data samples.
Applying these techniques to the full set of observed data, however, yields
a maximum at $\sigma = (0,0,0)$ which corresponds to the sum of delta functions
result.  To avoid this difficulty, maximum likelihood 
methods for bandwidth selection have 
been developed that are based on leave-one-out cross-validation~\cite{Sheather}.

The computational demands of leave-one-out cross-validation techniques
are excessive for our HITf/x data set.
Therefore, we have adopted a cross-validation 
method which requires less computation.
From the full set of $n$ observed $x_i$ vectors, we generate
$M$ disjoint subsets~$S_j$ of fixed size $n_v$ to be
used for validation.  For each validation set $S_j,$ we construct the estimate
$\widehat p(x)$ using the $n - n_v$ vectors that are not in $S_j$ as a function
of the bandwidth vector $\sigma = (\sigma_s,\sigma_v,\sigma_h).$ 
The optimal bandwidth vector 
$\sigma_j^* = (\sigma_{sj}^*,\sigma_{vj}^*,\sigma_{hj}^*)$ for $S_j$ is the 
choice that maximizes the pseudolikelihood 
\cite{Duin} \cite{keddR} according to 

\begin{equation}
\sigma_j^* = \arg\max_{_{\hspace{-0.8cm} \vspace{1cm} {\textstyle \sigma}}} \prod_{x_i \in S_j} \widehat p (x_i)
\label{eq:pseudo}
\end{equation}

\noindent where the product is over the $n_v$ vectors in the validation set $S_j.$
The overall optimized bandwidth vector $\sigma^*$ 
is obtained by averaging the $M$ vectors $\sigma_j^*.$  We will present specific 
details of our implementation of this cross-validation method in
section~\ref{sec-cv}.

\subsection{Constructing the Estimate for \boldmath{$P(R_j | x)$}}
\label{sec-cons}

An estimate for $P(R_j | x)$ can be derived from estimates of the
quantities on the right side of equation~(\ref{eq:bayes}).
The density estimate $\widehat p(x)$ for $p(x)$ is obtained using the kernel method
defined by equations~(\ref{eq:kde}) and
(\ref{eq:oneD}) with the optimized bandwidth vector $\sigma^*$ learned using the process
described in section~\ref{sec-bandsel}.  
Each conditional probability density function
$p(x | R_j)$  is estimated in the same way
except that the training set is defined by the subset of the $x_i$ vectors 
with outcome $R_j.$ 
Since reduced sample sizes for specific outcomes $R_j$ preclude the learning of
individual bandwidth vectors for each $p(x | R_j),$  we use the 
$\sigma^*$ derived for $p(x)$ for each
case.  This approach also
has the desirable effect of providing the same smoothing to a batted ball vector
in the numerator and denominator of (\ref{eq:bayes}) which prevents
a probability $P(R_j | x)$ from exceeding one.
Each prior probability $P(R_j)$ 
is estimated by the fraction of the $n$ batted balls in the full training set with
outcome
$R_j.$  The estimate for $P(R_j | x)$ is then constructed by combining the estimates for
$p(x | R_j), P(R_j),$  and  $p(x)$ according to Bayes theorem.

\subsection{Intrinsic Value using wOBA}

Our goal is to combine the posterior probabilities $P(R_j | x)$ into a measure of
the intrinsic value of a batted ball as a function of $x.$  
If we ignore bunts and treat sacrifice flies as ordinary outs, then 
the expected value
of the traditional baseball statistics batting average and slugging percentage
for a batted ball with parameter vector $x$
can be obtained from a linear combination of the $P(R_j | x)$ probabilities.
These statistics, however, are deficient for describing the value
of a batted ball.  Batting average, for example, gives a home run and a single the same value
while slugging percentage overvalues a home run at four times the value of a single.
In 2007, Tango and his collaborators \cite{Tangobook} defined wOBA
as a linear combination of the probabilities of events in a baseball game.  The wOBA
coefficients that define the linear combination are derived from the average run value of each
event.  The original wOBA formulation includes events such as strikeouts and walks, but for
our purposes we restrict the analysis to batted balls.  The resulting formula is

\begin{equation}
{\rm wOBA}(x) = \sum_{j=0}^5 w_j P(R_j | x)
\label{eq:woba}
\end{equation}

\noindent where the $w_j$ are the coefficients 
for the six batted ball outcomes $R_0=$ out, $R_1=$ single, $R_2=$ double, $R_3=$ triple, $R_4=$ home run,     
and $R_5=$ batter reaches on error.

\section{Data Analysis}
\label{sec-da}

\subsection{Batted Ball Data and wOBA Coefficients}
\label{sec-bbd}

The HITf/x data used for this study was provided by SportVision and includes measurements
from every regular-season MLB game during 2014.  The data set used for estimating
wOBA$(x)$ consists of all balls in play 
with a horizontal angle in fair territory $(h \in [45^\circ,135^\circ])$ that were
tracked by the system where bunts are excluded.  This results in a set of 
$n = 124364$ batted balls which represents more than ninety-seven percent of the MLB
total for 2014.  The weights $w_j$ in equation~(\ref{eq:woba})  depend
on the run environment and can change from year-to-year.  Thus, for this 
2014 batted ball data we use the coefficients
$w_0 = 0.000, w_1 = 0.892, w_2 = 1.283,  w_3 = 1.635, 
 w_4 = 2.135,$ and $w_5 = 0.920$ 
where $w_0, w_1, w_2, w_3,$ and $w_4$ for 2014 were obtained from~\cite{fangraphsguts} and
$w_5$ was obtained from~\cite{Tangobook}.
In the following subsections we present
details of the algorithm described in section~\ref{sec-learning}
for learning wOBA$(x)$ from this data .

\subsection{Cross-Validation for Density Estimation}
\label{sec-cv}

The Bayesian method described in section~\ref{sec-bayes} uses probability
density estimates to compute the posterior probabilities $P(R_j | x).$  We use
the model described in section~\ref{sec-kde} 
with the cross-validation method described in section~\ref{sec-bandsel} 
to estimate the densities. 
Five validation sets $S_1, S_2, S_3, S_4,$ and  $S_5$ were used to select 
the optimized bandwidth vector $\sigma^*$ for the $p(x)$ estimate. 
Set $S_i$ includes $n_v$ batted balls that were hit on day $6i - 5$ of a calendar
month.  Set $S_2$, for example, includes only batted balls hit on the 7th day of a month.
The size $n_v = 3820$ was taken to be the largest value so that each set $S_i$ includes
the same number of elements.  
The decision to use six days of separation for the validation sets was made with the goal of 
maximizing the independence of the sets.  A regular-season series of consecutive
games between the same pair of teams always lasts less than six days.  In addition, major
league teams in 2014 tended to use a rotation of starting pitchers that repeats every
five days so that, if this tendency is followed, each starting pitcher will 
occur once per calendar month in each of the five validation sets.

For each validation set $S_j,$ a three-dimensional search was conducted
with a step size of 0.1 in 
$\sigma_{s}, \sigma_{v},$ and $\sigma_{h}$ to find
the optimized $\sigma_j^*$ in equation~(\ref{eq:pseudo}). 
For each $S_j$ and $\sigma$ vector under consideration, 
we removed the twenty $x_i$ batted ball vectors with the smallest value of $\widehat p (x_i)$ to prevent 
outliers from influencing the optimization.  The vectors $\sigma_j^*$ for 
each $S_j$ are given in Table~\ref{tab-sigma} and after averaging yielded
an optimized 
$\sigma^* = (\sigma_{s}^*,\sigma_{v}^*,\sigma_{h}^*)$ of $(2.02,1.50,2.20).$ 
We see that vertical angle has the smallest smoothing parameter 
$(\sigma_{v}^* = 1.50)$ which is
consistent with the observation from figures~\ref{speedhist}  and 
\ref{vhhist} 
that vertical angle has more detailed structure in its density than
batted ball speed or horizontal angle.

\begin{table}[h]
\caption{Optimal bandwidths $\sigma_j^*$ for validation sets $S_j$}
\begin{center}
\begin{tabular} {|c|c|c|c|c|}
\hline
$S_1$ & $S_2$ & $S_3$ & $S_4$ & $S_5$ \\
\hline
 (2.0,1.5,2.2) & (1.9,1.5,2.3) & (2.0,1.6,2.0)  & (2.0,1.6,2.3)  & (2.2,1.3,2.2)  \\
\hline
\end{tabular}
\end{center}
\label{tab-sigma}
\end{table}

\subsection{Visualizing the wOBA\boldmath{$(x)$} Function}

\begin{figure*}[!t]
\begin{center}
\begin{tabular}{c}
\vspace*{0.3cm}
\hspace*{-0.0cm}\includegraphics[height=12.138cm,width=18.207cm]{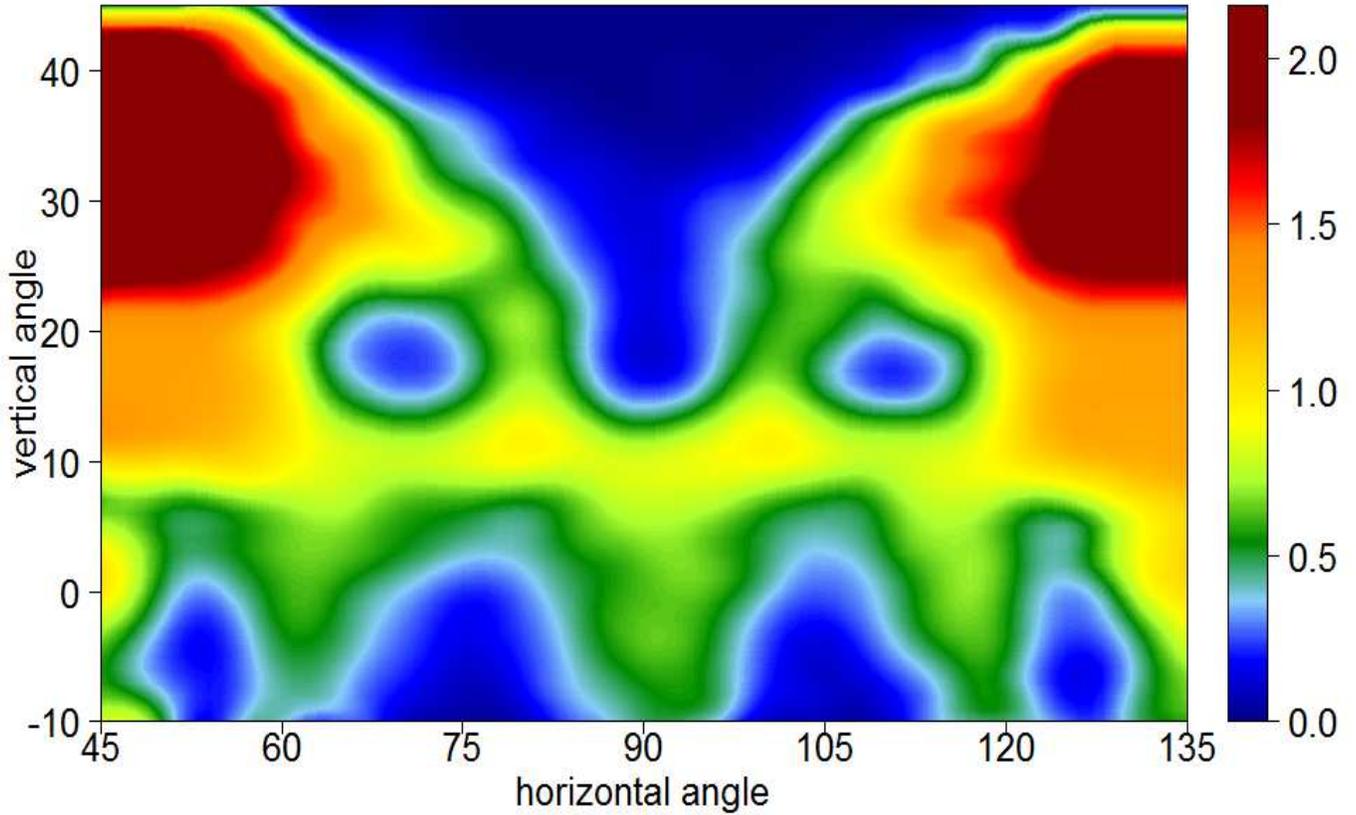}
\end{tabular}
\end{center}
\vspace*{-0.6cm}\caption{wOBA for an initial speed of 93 mph}
{ \label{heatmap22}
}
\end{figure*}

We constructed the function wOBA$(x)$ for 2014 HITf/x data using the
methods described in the previous sections.
Figure~\ref{heatmap22} displays wOBA$(x)$ on the plane corresponding to a 
fixed initial speed $s$ of ninety-three miles per hour.  For this value
of $s,$ the
best results for batters occur for balls hit with vertical angles between
twenty-five and forty degrees that are near the right field foul line
$(h \in [45^\circ,55^\circ])$ or the left field foul line
$(h \in [125^\circ,135^\circ])$ where ballpark dimensions are typically the
shortest.  These batted balls often result in home runs.  Batted balls hit
at the same speed with the same vertical angle are less valuable
at horizontal angles near ninety degrees which correspond to 
larger ballpark dimensions in center field.
For this initial speed, batted balls with vertical angles near 
twelve degrees tend to carry over the infielders and land in front of
the outfielders and have a high value for all horizontal angles.  Typical 
horizontal angle positions for the 
three outfielders are
evident from the three cold zones for balls hit in the air with
$(v \in [15^\circ,20^\circ])$ and typical horizontal positions for the
four infielders are evident from the four cold zones for ground balls
($v < 0).$

\begin{figure}[h]
\begin{center}
\begin{tabular}{c}
\vspace*{0.3cm}
\hspace*{-1.9cm}\includegraphics[height=7.48cm,width=11.22cm]{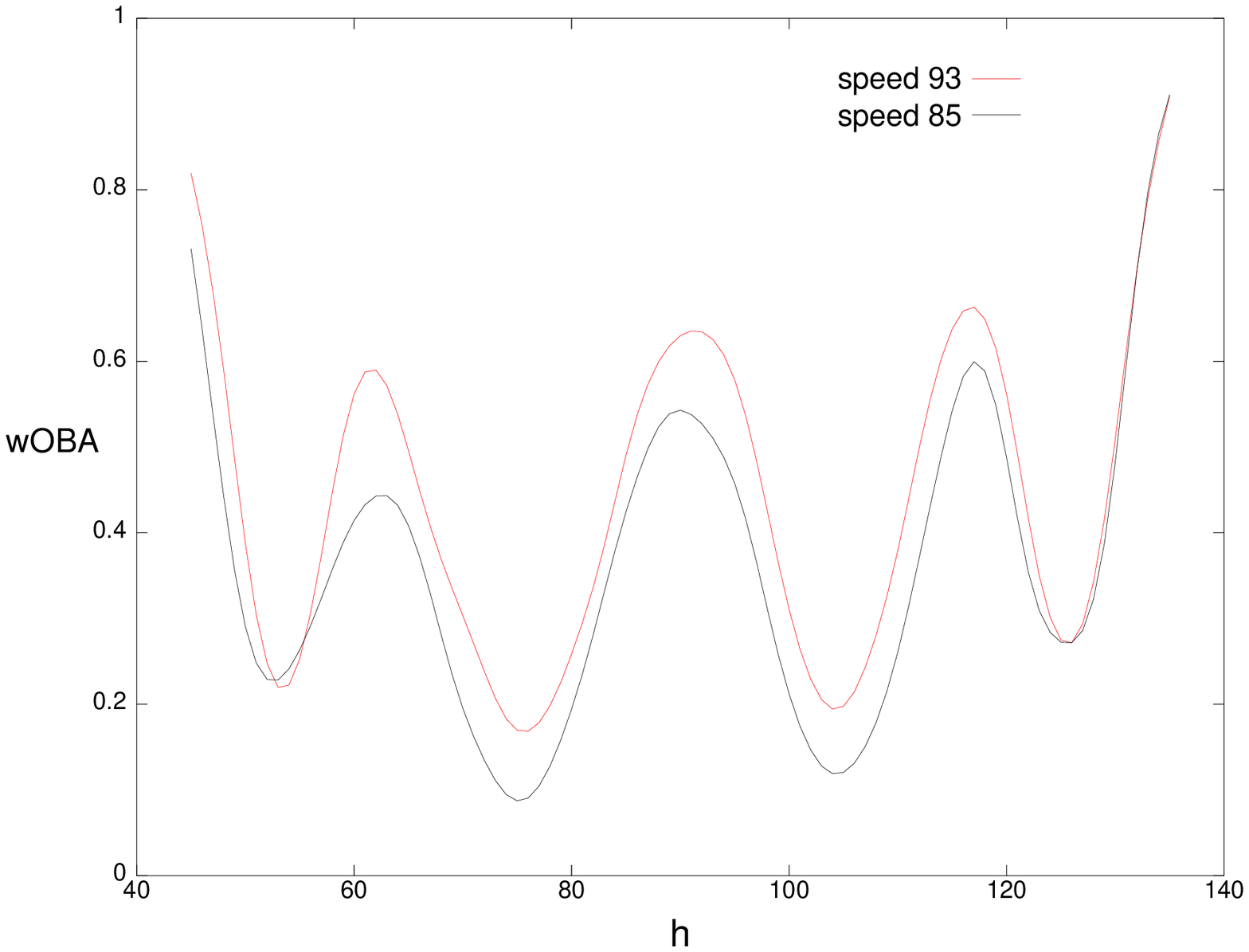}
\end{tabular}
\end{center}
\vspace*{-0.6cm}\caption{wOBA for a vertical angle $v$ of $-2^\circ$}
{ \label{fig3}
}
\end{figure}

\begin{figure}[h]
\begin{center}
\begin{tabular}{c}
\vspace*{0.3cm}
\hspace*{-0.9cm}\includegraphics[height=7.48cm,width=11.22cm]{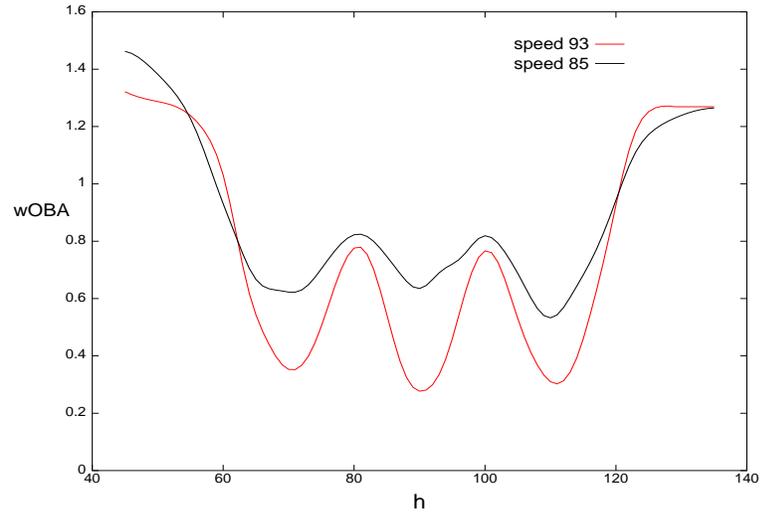}
\end{tabular}
\end{center}
\vspace*{-0.6cm}\caption{wOBA for a vertical angle $v$ of $+15^\circ$}
{ \label{fig4}
}
\end{figure}

Figures~\ref{fig3} and \ref{fig4} examine properties of the wOBA$(x)$ 
function in more detail.
Figure~\ref{fig3} plots wOBA as a function of the horizontal angle $h$
for $s=85$ mph and $s=93$ mph with the vertical angle fixed at $v=-2^\circ$.
Since this negative value of $v$ corresponds to ground balls,  minima 
in the two curves correspond to the typical position of infielders with
the minima near 
$53^\circ,76^\circ,104^\circ,$ and $126^\circ$ corresponding to the first baseman, 
second baseman, shortstop, and third baseman respectively.  
Over most horizontal angles, balls hit at 93 mph
have a higher value than balls hit at 85 mph since ground balls hit
at a higher speed have a higher probability of eluding a defender.
Figure~\ref{fig4} plots wOBA as a function of the horizontal angle $h$
for $s=85$ and $s=93$ with the vertical angle fixed at $v=+15^\circ$.
Since this positive value of $v$ corresponds to balls hit in the air,  
minima in the two curves correspond to the typical position of outfielders
with the minima near 
$70^\circ,90^\circ,$ and $110^\circ$ corresponding to the right fielder,
center fielder, and left fielder respectively.  
For this vertical angle, balls hit in the direction of an outfielder
have a higher value for a speed of 85 mph because these balls often
fall in front of the outfielder for hits while balls hit 
at 93 mph more frequently carry to the outfielder for outs.
In both figures, the largest
wOBA values occur for balls hit near the foul lines 
($h=45^\circ$ or $h=135^\circ$) which often result in extra-base hits
instead of singles.

\subsection{Dependence of wOBA\boldmath{$(x)$} on Batter Handedness}

Significant wOBA$(x)$ differences 
between left-handed and right-handed batters
occur due to differences in the positioning of defenders.  Thus, we 
repeated the process described in the previous sections
to obtain wOBA$l(x)$ for left-handed batters and 
wOBA$r(x)$ for right-handed batters.  The $n = 124364$ batted balls described in
section~\ref{sec-bbd} were first partitioned into the 54948 for left-handed batters
and 69416 for right-handed batters.  The method described in section~\ref{sec-cv}
was then used to build five validation sets for each case which resulted in a validation
set size $n_v$ of 1680 for wOBA$l(x)$ and 2190 for wOBA$r(x).$ 
The optimal bandwidth vectors $\sigma_j^*$ for each validation set and batter handedness
are given in Table~\ref{tab-sigmaLR}.  After averaging, we arrive at an optimized
$\sigma^* = (\sigma_{s}^*,\sigma_{v}^*,\sigma_{h}^*)$ of $(2.18,1.72,2.50)$ for 
wOBA$l(x)$ and  $(2.16,1.56,2.30)$ for wOBA$r(x).$ 
We note that, as seen in section~\ref{sec-cv}, $\sigma_{v}^*$ is the smallest for
each case while $\sigma_{h}^*$ is the largest.  In addition, the bandwidth increases for
each variable to provide more smoothing as the number of samples decreases.

\begin{table}[h]
\caption{Optimal bandwidths $\sigma_j^*$ for validation sets $S_j$ by batter handedness}
\begin{center}
\hspace{-0.9cm}
\begin{tabular} {|c|c|c|c|c|c|}
\hline
  &  $S_1$ & $S_2$ & $S_3$ & $S_4$ & $S_5$ \\
\hline
L &  (2.0,1.5,3.1) & (2.2,2.1,2.2) & (2.3,1.6,2.1)  & (2.4,1.9,2.3)  & (2.0,1.5,2.8)  \\
\hline
R & (1.9,1.8,2.1) & (2.1,1.7,2.2) & (2.4,1.4,2.2)  & (2.2,1.5,2.6)  & (2.2,1.4,2.4)  \\
\hline
\end{tabular}
\end{center}
\label{tab-sigmaLR}
\end{table}

\begin{figure}[h]
\begin{center}
\begin{tabular}{c}
\vspace*{0.3cm}
\hspace*{-0.7cm}\includegraphics[height=7.48cm,width=11.22cm]{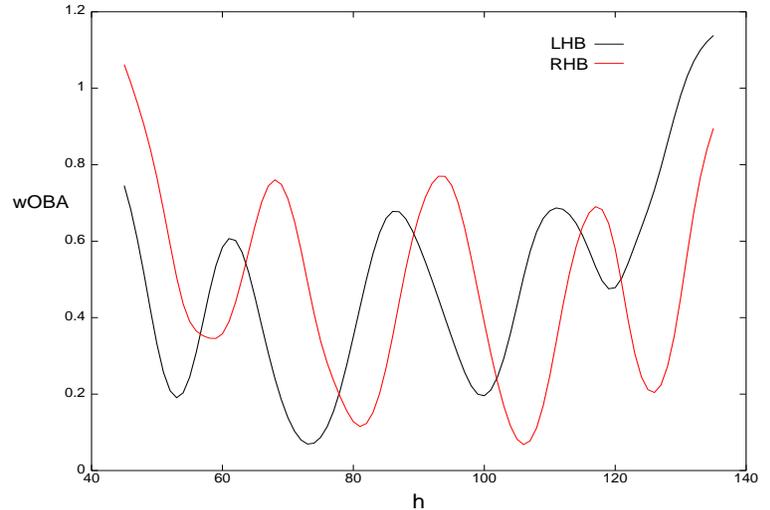}
\end{tabular}
\end{center}
\vspace*{-0.6cm}\caption{wOBA for speed 93 mph and vertical angle $-2^\circ$}
{ \label{fig5}
}
\end{figure}

\begin{figure}[h]
\begin{center}
\begin{tabular}{c}
\vspace*{0.3cm}
\hspace*{-0.7cm}\includegraphics[height=7.48cm,width=11.22cm]{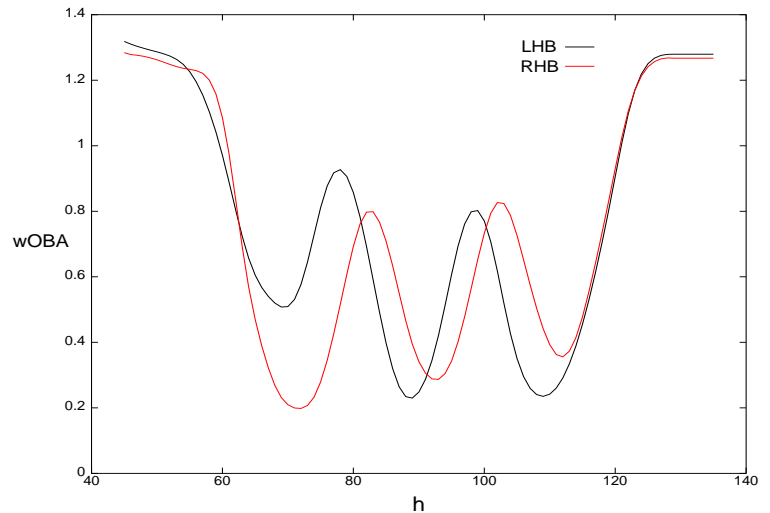}
\end{tabular}
\end{center}
\vspace*{-0.6cm}\caption{wOBA for speed 93 mph and vertical angle $+15^\circ$}
{ \label{fig6}
}
\end{figure}

Figures~\ref{fig5} and \ref{fig6} examine  differences between
wOBA$l(x)$  and wOBA$r(x).$   These figures consider specifically
batted balls hit at 93 miles per hour at 
the vertical angles shown in figure~\ref{fig3} and figure \ref{fig4}. 
Figure~\ref{fig5} considers ground balls
hit with a vertical angle of $-2^\circ.$  As in figure~\ref{fig3}, we observe four 
minima in each curve that correspond to the typical position of the four infielders.
We see, however, that the minima for left-handed batters are shifted several degrees
toward the first base line $(h = 45^\circ)$ compared to the corresponding minima for 
right-handed batters.  This shift corresponds to the difference in fielder positioning
as a function of batter handedness.  We also see that ground balls near the first base
line $(h = 45^\circ)$ have a higher value for right-handed batters since there is
a lower probability of a defender in that region and that ground balls near the third
base line $(h = 135^\circ)$ have a higher value for left-handed batters.
Figure~\ref{fig6} examines the impact of batter handedness on balls hit at
93 miles per hour with a vertical angle of $+15^\circ.$  The three minima in each curve
correspond to the typical positions of outfielders.  We see that
the minima are shifted several degrees toward the right field line $(h = 45^\circ)$ 
for left-handed batters.  We also see that left-handed batters have an advantage 
for batted balls hit in the direction of the right fielder ($h \approx 70^\circ$) since
the right fielder is typically playing deeper for left-handed batters which allows
additional batted balls to fall safely for hits.  We observe the opposite effect for
batted balls hit in the direction of the left fielder ($h \approx 110^\circ$)
since the left fielder is typically playing deeper for right handed batters.

\section{Exploiting the Model}

\subsection{Intrinsic Value versus Observed Outcome}
\label{sec-iv}

Using the model developed in sections \ref{sec-learning} and \ref{sec-da}, a batted ball $b$
with parameter vector $x$ can be assigned the intrinsic value $I(b)$ given by either 
wOBA$l(x)$ or wOBA$r(x)$ depending on the handedness of the batter.  Batted balls may also
be assigned an observed value $O(b)$ given by the wOBA coefficient for the result of the
batted ball so that, for example, $O(b) = 1.283$ for a batted ball that results in a double.  
The observed value $O(b)$ depends on several factors that are beyond the control of the batter
and the pitcher such as the defense, the weather, and the ballpark.  In early
May 2014, for example, batter Evan Longoria hit a high fly ball with an $I(b)$ of $0.040$
that is usually an easy out, but the center fielder lost the ball in the lights and the
result was a triple with an $O(b)$ of 1.635.
In this section, we describe several statistics
that can be generated using the intrinsic $I(b)$ value for a batted ball.

\subsection{Intrinsic Contact Measure for Batters}
\label{sec-ICMB}

Analysts sometimes attempt to quantify the value of a hitter's batted balls 
using the average 
$\overline O$ of his observed $O(b)$ over a period of time.
This statistic $\overline O$ is referred to as wOBA on contact or
wOBAcon \cite{Judge}.  As we pointed out in section~\ref{sec-iv}, however, $\overline O$
depends on a number of variables that are independent of the batter's quality of 
contact.   Thus, we propose the average $\overline I$ of the
intrinsic values $I(b)$ as a more accurate valuation of a hitter's collection of
batted balls.   Table~\ref{tab-batterquality} presents the batters with the
highest $\overline I$ in 2014 among players who hit at least 300 batted balls
that were tracked by HITf/x.  These players are known for their
ability to generate hard-hit balls.

\begin{table}[h]
\caption{Batters with the highest $\overline I$ in 2014 over at least 300 batted balls}
\begin{center}
\begin{tabular} {|c|c|}
\hline
\rule{0pt}{2.5ex} \hspace*{-.3cm} Batter &  $\overline I$ \\
\hline
Giancarlo Stanton & 0.526 \\
\hline
Mike Trout        & 0.498 \\
\hline
Miguel Cabrera    & 0.488 \\
\hline 
J.D. Martinez     & 0.482 \\
\hline
Matt Kemp         & 0.477 \\
\hline
Brandon Moss      & 0.476 \\
\hline
Jose Abreu        & 0.469 \\
\hline
Michael Morse     & 0.468 \\
\hline
Corey Dickerson   & 0.465 \\
\hline
Edwin Encarnacion & 0.461 \\
\hline
\end{tabular}
\end{center}
\label{tab-batterquality}
\end{table}

For an individual batter, several factors can contribute to differences between
the average observed outcome $\overline O$ and the intrinsic value $\overline I$
of his batted balls.  
Batters who are fast runners, for example, force infielders to play
shallower which compromises range and leads to additional hits.  Fast runners
also tend to beat out more infield hits and garner additional bases on hits to the
outfield.  Thus, a faster runner will tend to achieve a higher $\overline O$
for a given $\overline I.$
Batters with a high degree of predictability in their batted balls, such as
left-handed batters who hit a large majority of their ground balls to the
right of second base, are easier to defend than batters who produce a more
uniform distribution of batted balls.  Batters with a higher
degree of predictability, therefore, will tend to have a lower $\overline O$
for a given $\overline I.$   Random noise, which is often referred to as
luck in this context, can also play a role in 
creating differences between $\overline O$ and $\overline I.$  Thus,
our model for $\overline I$ is a useful starting point for 
separating the contribution of factors such as intrinsic contact ability, 
running speed, batted ball 
distribution, and luck on the observed $\overline O.$
Since noise contributions will tend to be independent from year-to-year and 
the other factors  will be represented by statistics with different reliabilities
and aging curves, the ability to separate these factors has 
value for projection systems.

Table~\ref{tab-OminusIhigh} presents batters with the largest values
of $\overline O - \overline I$ during the 2014 season where both 
$\overline O$ and $\overline I$ are computed using the batted balls 
tracked by HITf/x.   Most of the players in the table
have above average running speed with Hamilton and Cain having 
exceptional speed.   The top two players on the list also benefited from
good luck.  Starling Marte led major league baseball by reaching base on 
an error fourteen times in 2014 which contributed to his MLB leading
$\overline O - \overline I.$    Jose Abreu also experienced significant
good fortune as many of his 36 home runs just barely cleared the fence
causing his home runs to have an average intrinsic value
$I(b)$ of 1.461 which is significantly less than the corresponding 
$O(b)$ of 2.135.  Abreu's
$\overline O - \overline I$ difference on home runs explains 
nearly all of his total $\overline O - \overline I$ difference in the table and
his luck on home runs even attracted the attention of the
mainstream media during the 2014 season~\cite{wallstreetjournal}.

\begin{table}[h]
\caption{Batters with the highest $\overline O - \overline I$ in 2014 over at least 300 batted balls}
\begin{center}
\begin{tabular} {|c|c|}
\hline
\rule{0pt}{2.5ex} \hspace*{-.3cm} Batter &  $\overline O - \overline I$ \\
\hline
Starling Marte    & 0.072 \\
\hline
Jose Abreu        & 0.063 \\
\hline
Yasiel Puig       & 0.060 \\
\hline 
Adam Eaton        & 0.060 \\
\hline
Billy Hamilton    & 0.060 \\
\hline
Lorenzo Cain      & 0.059 \\
\hline
J.D. Martinez     & 0.053 \\
\hline
Josh Harrison     & 0.052 \\
\hline
Andrew McCutchen  & 0.051 \\
\hline
Hunter Pence      & 0.049 \\
\hline
\end{tabular}
\end{center}
\label{tab-OminusIhigh}
\end{table}

Table~\ref{tab-OminusIlow} presents batters with the lowest values
of $\overline O - \overline I$ during the 2014 season using 
the batted balls tracked by HITf/x.   All of the batters in this table
have below average running speed and several (Moss, Teixeira, Santana)
also had sufficiently predictable batted ball distributions that opposing teams
were able to employ extreme defensive shifts.  These factors contributed to
the small $\overline O - \overline I$  values shown in 
Table~\ref{tab-OminusIlow}. 

\begin{table}[h]
\caption{Batters with the Lowest $\overline O - \overline I$ in 2014 over at least 300 batted balls}
\begin{center}
\begin{tabular} {|c|c|}
\hline
\rule{0pt}{2.5ex} \hspace*{-.3cm} Batter &  $\overline O - \overline I$ \\
\hline
Billy Butler      & -0.055 \\
\hline
Brandon Moss      & -0.043 \\
\hline
Yadier Molina     & -0.042 \\
\hline 
Miguel Cabrera    & -0.042 \\
\hline
Matt Dominguez    & -0.041 \\
\hline
Alberto Callaspo  & -0.039 \\
\hline
Mark Teixeira     & -0.038 \\
\hline
Albert Pujols     & -0.038 \\
\hline
Carlos Santana    & -0.036 \\
\hline
Buster Posey      & -0.032 \\
\hline
\end{tabular}
\end{center}
\label{tab-OminusIlow}
\end{table}

\subsection{Intrinsic Opponent Contact Measure for Pitchers}
\label{sec-ioc}

In 2001, McCracken \cite{McCracken} suggested that pitchers have little
control over the result of opponent batted balls that are not home runs.
Since then, however, a number of researchers~\cite{Bradbury} \cite{Fast111611} 
\cite{Fast112211} 
\cite{Lichtman} \cite{Swartz121510} \cite{Swartzbathwater} \cite{Tippett} have presented 
evidence that pitchers can affect the expected outcome of balls in play.
Despite this progress, however, 
models that isolate the impact of the pitcher on the fate of batted balls
have been elusive due to the confounding effects of the defense, ballpark,
weather, and luck on a batted ball's outcome.
Since the HITf/x system characterizes a batted ball at contact, the influence of 
these confounding factors can be removed.
As proposed in section~\ref{sec-ICMB} for batters,
we can assign the intrinsic value $\overline I$ to the
collection of batted balls allowed by a pitcher.
The statistic $\overline I$ provides a context-invariant measure of 
a pitcher's opponent contact which allows this aspect of
his performance to be accurately quantified.

Table~\ref{tab-pitcherquality} presents the pitchers with the
lowest $\overline I$ values in 2014 among those who allowed at least 300 batted balls
that were tracked by HITf/x.
Eight of the ten pitchers in the table had an average fastball
speed in 2014 that was above the league average and Garrett Richards, who 
earned the top spot in Table~\ref{tab-pitcherquality}, enjoyed 
one of the highest average fastball speeds in MLB.  The success of the
two softer tossing pitchers on the list was due in part to an exceptional sinker
for Dallas Keuchel and an exceptional split-finger fastball for Alex Cobb.
An interesting topic for future research will be to study pitcher
characteristics that lead to low values of~$\overline I.$

\begin{table}[h]
\caption{Pitchers with the lowest opponent $\overline I$ over at least 300 batted balls, 2014}
\begin{center}
\begin{tabular} {|c|c|}
\hline
\rule{0pt}{2.5ex} \hspace*{-.3cm} Pitcher &  $\overline I$ \\
\hline
Garrett Richards  & 0.304 \\
\hline
Anibal Sanchez    & 0.309 \\
\hline
Danny Duffy       & 0.314 \\
\hline 
Chris Sale        & 0.319 \\
\hline
Matt Garza        & 0.328 \\
\hline
Dallas Keuchel    & 0.329 \\
\hline
Jarred Cosart     & 0.329 \\
\hline
Clayton Kershaw   & 0.332 \\
\hline
Alex Cobb         & 0.336 \\
\hline
Johnny Cueto      & 0.337 \\
\hline
\end{tabular}
\end{center}
\label{tab-pitcherquality}
\end{table}

\vspace{-0.4cm}

\subsection{Defense and Environmental Effects}

\subsubsection{Contact-Adjusted Defense}
\label{sec-cad}

Many statistics have been designed to measure team defense.  Defensive efficiency ratio (DER) \cite{james1978}, 
for example, measures the fraction of the time that a team's defense records an out on a batted
ball that is not a home run.  
While DER is a useful measure of defense that is easy to compute, the statistic does not
account for the difficulty of fielding a batted ball or distinguish between 
different results such as a single or a double.
These deficiencies have led to the development of the advanced fielding metrics defensive runs
saved (DRS) \cite{drs} and ultimate zone rating (UZR) \cite{uzr}.  These metrics use data from
Baseball Info Solutions (BIS) that partition batted balls into types such as bunts, ground balls,
line drives, and fly balls and the speed categories of soft, medium, and hard.  In addition, 
the DRS system uses video scouts and timer data to further partition batted balls into speed
bins that have a width of 10 mph so that, for example, batted balls hit 
between 65 and 75~mph are sorted into the same bin.  DRS and UZR generate measures of player and 
team defense in units of runs above average.

The use of HITf/x data 
has the potential to improve on the accuracy of defensive metrics by exploiting
higher resolution measurements of batted ball speed and direction than are available with
BIS data.  Using the 
difference between the intrinsic $I(b)$ and observed $O(b)$ value of a batted ball as defined 
in section~\ref{sec-iv}, we obtain a simple automated technique for measuring contact-adjusted 
team defense that can serve as the basis 
for a HITf/x-based defensive metric.  
For this application, we build the underlying wOBA$l(x)$ and wOBA$r(x)$ models using the subset
of batted balls that are not home runs.  For the 2014 HITf/x data this results in 120231 batted
balls.  Each team's contact-adjusted defense is defined as $D = \overline I - \overline O$ where
the averages $\overline I$ and $\overline O$ 
are computed over the batted balls tracked by HITf/x that are not home runs  
while the team is in the field.
Thus $D$ will be positive for teams that 
are above average and negative for teams that are below average.  Table~\ref{tab-CAD} presents
the teams with the top five and the bottom five values of $D$ among the 30 major league teams in 
2014.  We also include each team's corresponding rank according to the DRS and UZR systems.
We note that DRS and UZR include a number of additional factors which are not included
in $D$  such as bunt defense, the quality of outfield throwing arms, 
and the ability of infielders to turn double plays.  
Nevertheless, we see that the best defensive
teams according to $D$  tend to have high ratings according to DRS and
UZR and that the worst defensive teams according to $D$  tend to have low 
ratings according to DRS and UZR.  We note that Seattle receives a significantly
more favorable rating using $D$ relative to the other systems 
but this this can be explained by
environmental factors that will be discussed in section~\ref{sec-bpe}.

\begin{table}[h]
\caption{Teams with the highest and lowest Contact-Adjusted Defense $D,$ 2014}
\begin{center}
\begin{tabular} {|c|r|c|c|c|}
\hline
\rule{0pt}{2.5ex} \hspace*{-.3cm} Team &  $D$ \ \   & $D$ Rank & DRS Rank & UZR Rank \\
\hline
Oakland Athletics      & 0.018 &  1  & 7  & 8 \\
\hline
Baltimore Orioles        & 0.013 &  2  & 3  & 2 \\
\hline
Seattle Mariners       & 0.011 &  3  & 19 & 10 \\
\hline 
Cincinnati Reds           & 0.010 &  4  & 1  &  4 \\
\hline
San Diego Padres         & 0.009 &  5  & 5  &  9  \\
\hline
Toronto Blue Jays      & -0.013 & 26   & 23  & 21 \\
\hline
Chicago White Sox      & -0.017 & 27   & 27  & 26 \\
\hline
Detroit Tigers         & -0.018 & 28   & 28  & 28 \\
\hline
Minnesota Twins          & -0.018 & 29   & 29  & 25 \\
\hline
Cleveland Indians        & -0.018 & 30   & 30  & 30 \\
\hline
\end{tabular}
\end{center}
\label{tab-CAD}
\end{table}

\subsubsection{Environmental Effects}
\label{sec-bpe}

We can examine the impact of the environment on batted balls by comparing a
team's contact-adjusted defense for home and away games.  
For each team we define the difference 

\begin{equation}
D_{HA} = D_H - D_A
\end{equation}

\noindent where $D_H$ denotes $D$ for the team 
in home games and $D_A$ denotes $D$ for the team
in away games.  Thus, large positive values of 
$D_{HA}$  suggest that a team's home ballpark benefits the defense
while negative values suggest an environment that is less favorable for 
defenders.
Several factors can contribute to a team's observed $D_{HA}.$  
For a batted ball hit with 
a given parameter vector $x$ at contact, for example,  
the ambient wind and air density will have a significant 
effect on the speed of the ball as it travels 
which affects the ability of defenders to make a 
play~\cite{Adair}\cite{nathancolorado}.  
In addition, 
ballparks with slower infields or smaller outfield areas will tend to
be easier to defend.  Locations with high outfield walls lead to
smaller values of $D$ due to  
uncatchable balls that hit high on  
a wall but which are either easy outs or home runs in other parks.
We note that the DRS and UZR systems make 
adjustments for several of these environmental
factors.

Table~\ref{tab-CADHA} presents the major league
teams with the five highest and five lowest values of $D_{HA}$ 
during 2014.  We see that the environmental effects can be quite
substantial especially for teams with positive values of $D_{HA}.$ 
San Diego and Seattle have the largest values of
$D_{HA}$ which is consistent with observations of 
high air density in these locations.  In addition, 
Seattle, 
Philadelphia, Milwaukee, and Baltimore
play in home parks with smaller than average areas in the 
outfield~\cite{Gaines} which contributes to high values of $D_{HA}.$ 
There is also evidence that Philadelphia, Milwaukee, and Baltimore
had home parks with slow infields in 2014.  For games involving
the Orioles in 2014, there was a twenty-eight point higher batting
average on ground balls in road games than in home games and
the corresponding differences were fifteen points for Milwaukee and
ten points for Philadelphia.  Among teams with the smallest $D_{HA},$
Boston and Minnesota have home ballparks with high walls that 
lead to uncatchable balls in play.  The Rockies, Angels, and Mets
play in home ballparks with above average outfield areas which makes
batted balls more difficult to defend in these locations. 
In particular, the Rockies have
the ballpark with the largest outfield area while the Mets stadium
is the third largest.  In addition, games in Colorado are
characterized by a low air density \cite{nathancolorado}
which also tends to make defense more difficult.

\begin{table}[h]
\caption{Teams with the highest and lowest $D_{HA}$ in 2014}
\begin{center}
\begin{tabular} {|c|r|c|}
\hline
\rule{0pt}{2.5ex} \hspace*{-.3cm} Team &  $D_{HA}$ & Rank \\
\hline
San Diego Padres      & 0.043 &  1    \\
\hline
Seattle Mariners        & 0.031 &  2    \\
\hline
Philadelphia Phillies       & 0.026 &  3    \\
\hline 
Milwaukee Brewers           & 0.024 &  4    \\
\hline
Baltimore Orioles         & 0.021 &  5    \\
\hline
New York Mets      & -0.006 & 26     \\
\hline
Los Angeles Angels      & -0.009 & 27     \\
\hline
Colorado Rockies         & -0.014 & 28    \\
\hline
Minnesota Twins          & -0.017 & 29     \\
\hline
Boston Red Sox        & -0.018 & 30     \\
\hline
\end{tabular}
\end{center}
\label{tab-CADHA}
\end{table}

\section{Conclusion}

The amount of sensor data that is available to support sports analytics is
rapidly increasing~\cite{statcast} \cite{FastHBT} \cite{Jensen} \cite{nba}. 
This data has created unprecedented opportunities to exploit machine
learning techniques to model players and strategies
\cite{MillerICML} \cite{wei} \cite{YueICDM}.  
In this work we have used HITf/x
sensor data for more than one hundred thousand batted balls hit during the
2014 MLB season
to learn a model for a batted ball's intrinsic value which is invariant to
contextual factors
that can impact its outcome.  
The model is constructed 
using a Bayesian framework that includes kernel density estimates that are
learned using
cross-validation.  The intrinsic measure for contact quality is 
derived from the wOBA linear weights model for run value. 
The new approach successfully separates factors that are under the control of
the batter and pitcher from contextual factors that are characteristic
of the environment.

We have used the model developed in this paper to define statistics that
measure the intrinsic quality of contact for batters and pitchers.
These statistics are influenced less by random variation from contextual
variables than traditional statistics that depend on 
batted ball outcomes.  In addition, the new statistics can be used to separate 
the various skill 
components that contribute to a player's performance on batted balls.
A batter's performance, for example, can be partitioned into statistics
that measure
his intrinsic
contact, running speed, and batted ball distribution which determines
susceptibility to defensive shifts.  A pitcher's performance can be divided
into statistics that measure his opponent 
intrinsic contact and fielding ability.
An important advantage of generating separate statistics to represent
distinct skills is that
each statistic can be regressed and projected using its individual
reliability and aging curve during forecasting.
The new statistics also allow us to investigate 
how players control quality of contact.
In section~\ref{sec-ioc}, for example, we observed that many of the
pitchers who were the most effective at controlling contact also
exhibited an above-average fastball velocity.
Given the wealth of descriptors measured by the PITCHf/x 
system~\cite{FastHBT}, we have the opportunity to characterize
the relationship between the quality of 
a pitcher's opponent contact and his distribution and sequencing of pitches.
Similarly, we can study the relationship between a batter's intrinsic
contact and his swing parameters \cite{nathanswing}.

We have also examined the role of contextual factors that are beyond the
control of the batter and pitcher on the result of a batted ball.
In section~\ref{sec-cad} we used the intrinsic contact model developed in
this paper to derive a simple automated measure of contact-adjusted defense.
We showed that this measure is reasonably 
consistent with advanced defensive metrics
and that there are significant differences in defensive capability across
teams.  In 
section~\ref{sec-bpe} we showed that environmental factors such as ballpark
dimensions and weather conditions also have a significant effect on the
outcome of a batted ball.  Given the granularity of HITf/x data and the 
capability 
of density estimation techniques, 
the methods used in this paper could be adopted to design
defensive metrics and ballpark models that
are a function of the batted ball vector $x = (s,v,h).$  This would allow
a player's collection of batted balls to be translated to a new environment.
A forecasting system equipped with these models could accurately predict,
for example, 
how a given batter might perform in a new ballpark or how a given pitcher
might benefit from an improved infield defense.

% use section* for acknowledgment
\ifCLASSOPTIONcompsoc
  % The Computer Society usually uses the plural form
  \section*{Acknowledgments}
\else
  % regular IEEE prefers the singular form
  \section*{Acknowledgment}
\fi

I thank Josh Spivak and Graham Goldbeck at Sportvision for providing the
HITf/x data which made this work possible.  I am grateful to Qi Shi for 
his help with the preparation of this manuscript.

% Can use something like this to put references on a page
% by themselves when using endfloat and the captionsoff option.
\ifCLASSOPTIONcaptionsoff
  \newpage
\fi

\begin{IEEEbiographynophoto}{Glenn Healey}
is Professor of Electrical Engineering and Computer Science 
at the University of California, Irvine.  Before joining UC Irvine,
he worked at IBM Research.  Dr.~Healey
received the B.S.E. degree in Computer Engineering from
the University of Michigan and the M.S. degree in computer science, the M.S.
degree in mathematics, and the Ph.D. degree in computer science 
from Stanford University.  He is director of the Computer Vision
Laboratory at UC Irvine.  
Dr. Healey has served on the editorial boards of  IEEE Transactions on Pattern 
Analysis and Machine Intelligence,  IEEE Transactions on 
Image Processing, and the  Journal of the Optical Society of America A. 
He has been elected a Fellow of IEEE and SPIE. Dr. Healey 
has received several awards for outstanding  teaching and research.
\end{IEEEbiographynophoto}

\end{document}